\documentclass[a4paper,11pt]{article}
\usepackage{pos}
\usepackage{mathtools}
\usepackage{bm}

\newcommand{\Lag}{\mathcal{L}}
\newcommand{\Tr}{\mathrm{Tr}}
\newcommand{\St}{\mathrm{St}}

\newcommand{\QCD}{\mathrm{QCD}}
\newcommand{\EW}{\mathrm{EW}}
\newcommand{\GeV}{\mathrm{GeV}}

\title{Gauging Axionic Symmetries and Dark Matter: In memory of George Lazarides}
\ShortTitle{Gauging Axionic Symmetries and Dark Matter}

\author*[a]{Claudio Corian\`o}

\affiliation[a]{Dipartimento di Matematica e Fisica ``Ennio De Giorgi'', Universit\`a del Salento and INFN Sezione di Lecce,\\
Via per Arnesano, 73100 Lecce, Italy}

\emailAdd{claudio.coriano@unisalento.it}

\FullConference{Corfu Summer Institute 2025\\
Corfu, Greece\\
Session in memory of George Lazarides}

\abstract{%
These notes are written for a memorial Session dedicated to George Lazarides.
They revisit a joint work on the cosmology of a gauged axion and place it in a
broader line of ideas connecting anomalous gauge symmetries, orientifold
effective actions, Stueckelberg fields and dark matter.  In models with an
anomalous extra $U(1)$ symmetry, the Stueckelberg pseudoscalar participates in
the restoration of gauge invariance through Wess-Zumino counterterms and, after
electroweak symmetry breaking, may leave a physical axion-like state.  Its
cosmological history differs from that of an ordinary Peccei-Quinn axion: the
physical field appears only after Higgs-Stueckelberg mixing, is subject to
sequential electroweak and QCD misalignment, and can give an appreciable
dark-matter relic abundance only when the Stueckelberg scale is sufficiently
large.  This perspective connects naturally with George's earlier insight that
the vacuum structure of axion models must be understood together with the gauge
structure in which it is embedded. I dedicate these notes to his memory, with gratitude for
the collaboration and for the clarity with which he connected particle physics
to the early universe.
}

\begin{document}
\maketitle

\section{Introduction}

The line of work that eventually led to our paper with George Lazarides on a
gauged axion~\cite{Coriano:2010zz} did not start from a conventional
Peccei-Quinn model.  Its immediate origin was the effective description of
low-scale orientifold vacua developed with Nikos Irges and Elias
Kiritsis~\cite{Coriano:2005own}.  In those models the low-energy spectrum
contains, besides the Standard Model fields, extra anomalous $U(1)$ gauge
bosons, Ramond-Ramond axions, Stueckelberg mass terms and Wess-Zumino
interactions.  This structure was not an optional decoration of the effective
theory.  It was the four-dimensional remnant of anomaly cancellation in
orientifold compactifications and it controlled the neutral-current sector
after electroweak symmetry breaking.

The broader theoretical setting had been clarified in a series of works on
anomalous $U(1)$ symmetries in type-I and orientifold vacua
\cite{Antoniadis:2002cs,Kiritsis:2005qk}.  The analysis of anomaly-related
effective couplings, including Stueckelberg, axionic and generalized
Chern-Simons terms, was developed further in
\cite{Anastasopoulos:2006cz,Anastasopoulos:2007}.  These works made clear that
anomalous abelian factors could leave characteristic low-energy signatures even
after their gauge bosons became massive.  For us the question then became more
specific: if the Stueckelberg axions of such effective theories mix with the
CP-odd Higgs sector, can a physical axion-like particle remain in the spectrum,
and can it have a cosmological role?

This question was pursued first in field-theory language in
\cite{Coriano:2006xh}.  The decisive step was then developed with Nikos Irges
and Simone Morelli in the analysis of Stueckelberg axions and anomalous
abelian effective actions~\cite{Coriano:2007fw,Coriano:2007xg}.  There the
Higgs-axion mixing was studied from the viewpoint of unitarity, and the
$SU(3)_C\times SU(2)_W\times U(1)_Y\times U(1)_B$ realization made explicit
the anomalous neutral-current sector and its possible collider signatures.  The
physical pseudoscalar that emerges after Higgs-Stueckelberg mixing was called
the \emph{axi-Higgs}.  It is not a standard invisible axion.  Its interactions
are fixed by the anomaly-cancelling Wess-Zumino terms, and its mass is
controlled by non-perturbative contributions to the scalar potential.  In this
way the orientifold-inspired anomalous $U(1)$ construction suggested a new path
toward axion-like physics: the shift symmetry was gauged, the anomaly was
cancelled locally, and the physical pseudoscalar appeared only after symmetry
breaking.

This was the point at which George's perspective became especially natural.
George had always treated particle-physics models as cosmological systems.  A
model was not finished once its field content and symmetries were written down;
one also had to understand its vacuum structure, defects and relics.  In the
case of axions this attitude goes back to the Lazarides-Shafi mechanism, which
showed that the domain-wall problem can be avoided when the residual axionic
discrete symmetry is embedded in a continuous gauge group~\cite{Lazarides:1982tw}.
Thus the project with George was not simply to compute the relic density of
another light pseudoscalar.  It was to ask how the cosmology changes when the
axionic symmetry itself is gauged.

I had the opportunity to meet George in Crete, in Thessaloniki and in Lecce,
and on each occasion I was struck by the same quality: a wide scientific
perspective joined to a very direct and generous way of discussing physics.  He
moved naturally between grand unification, supersymmetric model building,
topological defects, inflation, axions, monopoles, cosmic strings and dark
matter.  These subjects were not separate compartments in his thinking.  They
were parts of one question: how can a microscopic theory of fundamental
interactions be made consistent with the history of the early universe?

\section{George Lazarides and the vacuum structure of axion models}

The Peccei-Quinn mechanism~\cite{Peccei:1977hh,Peccei:1977ur} remains one of
the most elegant solutions of the strong CP problem.  Its low-energy
consequence is a pseudo-Nambu-Goldstone boson, the axion
\cite{Weinberg:1977ma,Wilczek:1977pj}, whose mass and couplings are governed
by a high scale $f_a$.  The same field is also a compelling dark-matter
candidate.  The cosmological abundance produced by misalignment is controlled,
to first approximation, by the initial angle and by the scale $f_a$.

The standard story also contains a danger.  When the Peccei-Quinn symmetry is
spontaneously broken and QCD instantons generate an axion potential, different
vacua can remain separated by domain walls.  If the domain-wall number is
larger than one and the walls are stable, the resulting network is
cosmologically unacceptable~\cite{Sikivie:1982qv}.  George's contribution with
Qaisar Shafi was to show that the apparent multiplicity of vacua can be
removed in grand unified models if the discrete symmetry left by QCD is
embedded in the center of a continuous gauge group~\cite{Lazarides:1982tw}.  In
that case vacua that look distinct from the viewpoint of the low-energy global
symmetry are actually gauge equivalent.

This mechanism is usually remembered as a solution to the domain-wall problem,
but it also carries a more general lesson.  Axion physics depends not only on
the local form of the effective potential, but also on the global
identification of vacua.  The same philosophy appears in George's work on
strings and walls in grand unified theories~\cite{Kibble:1982ae} and in early
studies of axions as dark matter in Spin$(10)$ models~\cite{Holman:1982tb}.
It also belongs to a broader program in which grand unified theories,
supersymmetry and cosmology are tested against the production of monopoles,
strings, walls and relic particles.  George's work helped make this interplay
part of the standard language of particle cosmology.  A particularly important
part of this program was his work on supersymmetric hybrid inflation.  In the
smooth and shifted variants of hybrid inflation, the breaking of a grand
unified gauge symmetry can take place along the inflationary trajectory, so
that dangerous monopoles are diluted or never produced at the end of inflation
\cite{Lazarides:1995vr,Dvali:1997uq,Jeannerot:2000sv}.  This is again the same
kind of physics: a model of fundamental interactions is judged together with
its inflationary phase, its reheating history, its baryogenesis mechanism and
its topological relics.  For the subject of gauged axions, this inheritance is
conceptual.  An axion is not only a light pseudoscalar with a periodic
potential.  It is also a statement about symmetry, anomaly, vacuum
identification and cosmological history.

\section{The gauged axion setup}

The model discussed in Ref.~\cite{Coriano:2010zz} belongs to the class of
effective theories in which the gauge group of the Standard Model is enlarged
by an anomalous abelian factor, here denoted by $U(1)_B$.  The corresponding
gauge boson $B_\mu$ acquires a Stueckelberg mass through
\begin{equation}
  \Lag_{\St}
  =
  \frac{1}{2}\left(\partial_\mu b - M B_\mu\right)^2 ,
  \label{eq:stueckelberg}
\end{equation}
where $b$ is a pseudoscalar Stueckelberg field.  The local shift symmetry is
\begin{equation}
  B_\mu \to B_\mu + \partial_\mu \theta_B,
  \qquad
  b \to b + M\theta_B .
  \label{eq:local-shift}
\end{equation}
Thus $b$ is not, by itself, a physical axion in the high-energy Stueckelberg
phase.  It can be absorbed into the longitudinal component of the anomalous
gauge boson.

The anomalous fermion triangle diagrams are compensated by Wess-Zumino
counterterms and, in the mixed abelian and non-abelian sectors, by generalized
Chern-Simons terms.  Schematically,
\begin{equation}
  \Lag_{\mathrm{WZ}}
  =
  \frac{b}{M}
  \left[
      C_{BB} F_B\widetilde F_B
    + C_{YY} F_Y\widetilde F_Y
    + C_{WW} \Tr W\widetilde W
    + C_{GG} \Tr G\widetilde G
  \right],
  \label{eq:wz}
\end{equation}
with coefficients fixed by the anomaly cancellation conditions of the
effective action.  The important point is that the axion-like interactions are
not added by hand as global Peccei-Quinn interactions.  They are required by
the restoration of gauge invariance in the presence of the anomalous
$U(1)_B$.

At this stage the analogy with the Peccei-Quinn axion is incomplete.  In the
ordinary invisible axion model, a physical angular mode exists below the
Peccei-Quinn breaking scale.  In the gauged construction, the Stueckelberg
field becomes physically visible only after its mixing with the CP-odd sector
of the Higgs fields.

\section{The physical axi-Higgs}

The low-energy model contains two Higgs doublets, denoted by $H_u$ and $H_d$,
with vacuum expectation values $v_u$ and $v_d$.  After electroweak symmetry
breaking, the CP-odd sector contains the phases of the neutral Higgs fields and
the Stueckelberg field $b$.  Two combinations are Goldstone modes associated
with the neutral massive gauge bosons, while one physical pseudoscalar remains.
In the notation of Ref.~\cite{Coriano:2010zz}, this physical state was called
the axi-Higgs and denoted by $\chi$.

Gauge invariance allows non-perturbative operators involving the Higgs bilinear
and the Stueckelberg phase.  In schematic form, the relevant phase is
\begin{equation}
  \Theta
  =
  \frac{g_B(q_d-q_u)}{2M}\,b
  - \frac{\mathrm{Im}\,H_u^0}{v_u}
  + \frac{\mathrm{Im}\,H_d^0}{v_d}
  =
  \frac{\chi}{\sigma_\chi},
  \label{eq:theta}
\end{equation}
where $q_u$ and $q_d$ are the $U(1)_B$ charges of the two Higgs doublets.  The
normalization scale of the physical angle is
\begin{equation}
  \sigma_\chi
  =
  \frac{2v_u v_d M}
  {\left[g_B^2(q_d-q_u)^2v_u^2v_d^2
  +2M^2(v_u^2+v_d^2)\right]^{1/2}} .
  \label{eq:sigma}
\end{equation}
For a Stueckelberg scale well above the electroweak scale, $\sigma_\chi$ is of
order $v=\sqrt{v_u^2+v_d^2}$, not of order $M$.  This observation is central
for the cosmology of the model.

The non-perturbative potential can be written, at the level relevant for the
cosmological discussion, as a periodic potential for $\chi$,
\begin{equation}
  V_{\mathrm{np}}(\chi)
  =
  A\cos\left(\frac{\chi}{\sigma_\chi}\right)
  +
  B\cos\left(2\frac{\chi}{\sigma_\chi}\right),
  \label{eq:periodic-potential}
\end{equation}
where $A$ and $B$ are determined by the coefficients of the gauge-invariant
Higgs-Stueckelberg operators.  If this potential is generated at the
electroweak transition, its natural size is strongly suppressed by the weak
instanton action.  The mass induced for $\chi$ is therefore extremely small.

\section{Sequential misalignment and dark matter}

The cosmological behavior of $\chi$ differs from the usual Peccei-Quinn axion
in two essential ways.  First, before electroweak symmetry breaking, the
Stueckelberg field is not a physical scalar degree of freedom.  Second, the
field can experience two distinct misalignment events: one associated with the
electroweak transition, and a later one associated with QCD.

At the electroweak transition the relevant angular variable is
\begin{equation}
  \theta_{\EW} = \frac{\chi}{\sigma_\chi}.
\end{equation}
Since $\sigma_\chi$ remains of electroweak size, the relic abundance produced
by this first misalignment is very small for the parameter region studied in
Ref.~\cite{Coriano:2010zz}.  In that sense the physical axi-Higgs is present
after electroweak symmetry breaking, but it is almost frozen.

The more important contribution can arise at the QCD phase transition when the
mixed $U(1)_B$--$SU(3)$ anomaly generates a QCD potential.  At this point the
effective scale controlling the coupling to gluons is not $\sigma_\chi$ but
\begin{equation}
  f_{\chi}^{\QCD} \sim \frac{M^2}{v}.
  \label{eq:effective-decay}
\end{equation}
The QCD misalignment angle is therefore
\begin{equation}
  \theta_{\QCD} \simeq \frac{\chi}{f_{\chi}^{\QCD}}
  =
  \frac{\chi v}{M^2}.
  \label{eq:qcd-angle}
\end{equation}
The analogy with the invisible axion becomes transparent after the replacement
$f_a\to M^2/v$.  Correspondingly,
\begin{equation}
  m_\chi \sim \frac{\Lambda_{\QCD}^2}{f_{\chi}^{\QCD}}
  \sim \frac{\Lambda_{\QCD}^2 v}{M^2}.
  \label{eq:qcd-mass}
\end{equation}
This is the scale that controls the onset of coherent oscillations and the
late relic abundance.

The conclusion of the numerical analysis in Ref.~\cite{Coriano:2010zz} was
sharp.  For a Stueckelberg mass in the TeV range, the contribution of the
gauged axion to the present dark-matter abundance is negligible.  A sizeable
contribution appears only for much larger values of $M$, around an intermediate
scale, roughly $M\sim 10^7\,\GeV$, for which $M^2/v$ becomes comparable to the
usual invisible-axion scale.  Later discussions of Stueckelberg axions as dark
matter placed this result in a broader set of possibilities, including
intermediate and very high Stueckelberg scales~\cite{Coriano:2019}.

\section{Comments}

Looking back, what I associate most strongly with George in this subject is
the insistence that axion physics cannot be separated from the global structure
of the theory.  In the standard axion case, the question is whether apparently
distinct vacua are really distinct or are related by a gauge identification.
In the gauged axion case, the question is when the Stueckelberg field is a
physical particle and when it is a gauge artifact.  In both cases, the
cosmology follows from the answer. Our work was an attempt
to follow the fate of an anomalous gauge degree of freedom from the
high-energy Stueckelberg phase, through electroweak symmetry breaking, down to
the QCD epoch.  This path is precisely where George's intuition was most
valuable.  He was attentive to the places where a formal model can fail
cosmologically: unwanted walls, wrong vacuum multiplicities, dangerous relics,
or hidden assumptions about when a field becomes physical.

The phrase ``gauging axionic symmetries'' may sound technical, but in this
context it carries a simple lesson.  A local symmetry can protect the
consistency of an anomalous effective action, can reorganize the degrees of
freedom, and can change the cosmological initial conditions of the axion-like
state.  Dark matter is then not determined by a single mass and coupling, but
by the full history of symmetry realization.

\section{Conclusions}

George Lazarides's work left a deep imprint on the way we think about axions
in cosmology.  The Lazarides-Shafi mechanism showed that the domain-wall
problem can be solved by embedding the residual axionic discrete symmetry into
a continuous gauge symmetry.  Our later work on gauged axions explored a
different but related realization of the same broad idea: axionic degrees of
freedom should be understood together with the gauge symmetries and anomalies
from which they arise.

In the anomalous $U(1)$ model, the physical axion-like particle appears after
Higgs-\-Stueckelberg mixing.  Its electroweak misalignment gives a negligible
relic abundance, while its QCD misalignment can contribute to dark matter only
when the Stueckelberg scale is large enough that $M^2/v$ plays the role of an
effective axion decay constant.  This result is technically specific, but the
larger message is one I learned to associate with George: cosmology remembers
the symmetry structure of the theory.

There is also a human legacy that belongs in any remembrance of him.  George
was a great teacher and tutor for students.  He had the rare ability to make
young researchers feel that difficult questions were accessible if approached
with patience and clarity.  One sees this also in the work of younger
collaborators such as Roberta Armillis, whose paper with George and Constantinos
Pallis on inflation, leptogenesis and Yukawa quasiunification developed a
supersymmetric left-right framework in the same cosmological spirit
\cite{Armillis:2013wya}.  I also remember this influence in relation to Antonio
Mariano.  George's ideas on gauged axions, supersymmetry and cosmology were
instrumental for the later analysis of relic densities in the
$U(1)$-extended NMSSM and the gauged axion supermultiplet
\cite{Coriano:2010ox}.  His interactions with students and collaborators were
marked by kindness, attention and intellectual generosity.  For many people
this was as important as the scientific results themselves: he created an
atmosphere in which serious work could be done without losing the human warmth
that makes collaboration meaningful.

\section*{Acknowledgements}
I thank George Zoupanos and Margarida Nesbitt Rebelo for the invitation.


\begin{thebibliography}{99}

\bibitem{Coriano:2010zz}
C.~Corian\`o, M.~Guzzi, G.~Lazarides and A.~Mariano,
``Cosmological Properties of a Gauged Axion,''
Phys. Rev. D \textbf{82} (2010) 065013,
arXiv:1005.5441.

\bibitem{Coriano:2005own}
C.~Corian\`o, N.~Irges and E.~Kiritsis,
``On the Effective Theory of Low Scale Orientifold String Vacua,''
Nucl. Phys. B \textbf{746} (2006) 77,
arXiv:hep-ph/0510332.

\bibitem{Antoniadis:2002cs}
I.~Antoniadis, E.~Kiritsis and J.~Rizos,
``Anomalous U(1)'s in type I string vacua,''
Nucl. Phys. B \textbf{637} (2002) 92,
arXiv:hep-th/0204153.

\bibitem{Kiritsis:2005qk}
E.~Kiritsis,
``D-branes in standard model building, gravity and cosmology,''
Phys. Rept. \textbf{421} (2005) 105,
arXiv:hep-th/0310001.

\bibitem{Anastasopoulos:2006cz}
P.~Anastasopoulos, M.~Bianchi, E.~Dudas and E.~Kiritsis,
``Anomalies, anomalous U(1)'s and generalized Chern-Simons terms,''
JHEP \textbf{11} (2006) 057,
arXiv:hep-th/0605225.

\bibitem{Anastasopoulos:2007}
P.~Anastasopoulos,
``Anomalous U(1)'s, Chern-Simons couplings and the Standard Model,''
Fortsch. Phys. \textbf{55} (2007) 633,
arXiv:hep-th/0701114.

\bibitem{Coriano:2006xh}
C.~Corian\`o and N.~Irges,
``Windows over a New Low Energy Axion,''
Phys. Lett. B \textbf{651} (2007) 298,
arXiv:hep-ph/0612140.

\bibitem{Coriano:2007fw}
C.~Corian\`o, N.~Irges and S.~Morelli,
``Stueckelberg Axions and the Effective Action of Anomalous Abelian Models. 1.
A Unitarity Analysis of the Higgs-Axion Mixing,''
JHEP \textbf{07} (2007) 008,
arXiv:hep-ph/0701010.

\bibitem{Coriano:2007xg}
C.~Corian\`o, N.~Irges and S.~Morelli,
``Stueckelberg Axions and the Effective Action of Anomalous Abelian Models. II:
A $SU(3)_C\times SU(2)_W\times U(1)_Y\times U(1)_B$ Model and its Signature at
the LHC,''
Nucl. Phys. B \textbf{789} (2008) 133,
arXiv:hep-ph/0703127.

\bibitem{Lazarides:1982tw}
G.~Lazarides and Q.~Shafi,
``Axion Models with No Domain Wall Problem,''
Phys. Lett. B \textbf{115} (1982) 21.

\bibitem{Peccei:1977hh}
R.~D. Peccei and H.~R. Quinn,
``CP Conservation in the Presence of Instantons,''
Phys. Rev. Lett. \textbf{38} (1977) 1440.

\bibitem{Peccei:1977ur}
R.~D. Peccei and H.~R. Quinn,
``Constraints Imposed by CP Conservation in the Presence of Instantons,''
Phys. Rev. D \textbf{16} (1977) 1791.

\bibitem{Weinberg:1977ma}
S.~Weinberg,
``A New Light Boson?''
Phys. Rev. Lett. \textbf{40} (1978) 223.

\bibitem{Wilczek:1977pj}
F.~Wilczek,
``Problem of Strong P and T Invariance in the Presence of Instantons,''
Phys. Rev. Lett. \textbf{40} (1978) 279.

\bibitem{Sikivie:1982qv}
P.~Sikivie,
``Of Axions, Domain Walls and the Early Universe,''
Phys. Rev. Lett. \textbf{48} (1982) 1156.

\bibitem{Kibble:1982ae}
T.~W.~B. Kibble, G.~Lazarides and Q.~Shafi,
``Walls Bounded by Strings,''
Phys. Rev. D \textbf{26} (1982) 435.

\bibitem{Holman:1982tb}
R.~Holman, G.~Lazarides and Q.~Shafi,
``Axions and the Dark Matter of the Universe,''
Phys. Rev. D \textbf{27} (1983) 995.

\bibitem{Lazarides:1995vr}
G.~Lazarides and C.~Panagiotakopoulos,
``Smooth Hybrid Inflation,''
Phys. Rev. D \textbf{52} (1995) R559,
arXiv:hep-ph/9506325.

\bibitem{Dvali:1997uq}
G.~R. Dvali, G.~Lazarides and Q.~Shafi,
``Mu Problem and Hybrid Inflation in Supersymmetric
$SU(2)_L\times SU(2)_R\times U(1)_{B-L}$,''
Phys. Lett. B \textbf{424} (1998) 259,
arXiv:hep-ph/9710314.

\bibitem{Jeannerot:2000sv}
R.~Jeannerot, S.~Khalil, G.~Lazarides and Q.~Shafi,
``Inflation and Monopoles in Supersymmetric
$SU(4)_c\times SU(2)_L\times SU(2)_R$,''
JHEP \textbf{10} (2000) 012,
arXiv:hep-ph/0002151.

\bibitem{Coriano:2019}
C.~Corian\`o, P.~H. Frampton, N.~Irges and A.~Tatullo,
``Dark Matter With Stueckelberg Axions,''
Front. Phys. \textbf{7} (2019) 36,
arXiv:1811.05792.

\bibitem{Armillis:2013wya}
R.~Armillis, G.~Lazarides and C.~Pallis,
``Inflation, Leptogenesis, and Yukawa Quasiunification within a Supersymmetric
Left-Right Model,''
Phys. Rev. D \textbf{89} (2014) 065032,
arXiv:1309.6986.

\bibitem{Coriano:2010ox}
C.~Corian\`o, M.~Guzzi and A.~Mariano,
``Relic Densities of Dark Matter in the $U(1)$-Extended NMSSM and the Gauged
Axion Supermultiplet,''
Phys. Rev. D \textbf{85} (2012) 095008,
arXiv:1010.2010.

\end{thebibliography}
\end{document}